\documentclass[]{article}

\usepackage[utf8]{inputenc}

\usepackage{fullpage}
\usepackage{hyperref}
\usepackage{graphicx}

\usepackage{amsmath}

\usepackage{boxedminipage}

\usepackage{listings}
\usepackage{color}
\usepackage{textcomp}

\title{Python - All a Scientist Needs}

\author{Julius B. Lucks}

\date{\today}

\begin{document}

\maketitle

\begin{abstract}
    Any cutting-edge scientific research project requires a myriad of computational tools for data generation, management, analysis and visualization. Python is a flexible and extensible scientific programming platform that offered the perfect solution in our recent comparative genomics investigation \cite{LucksJB-PlOSCompBio-2008}. In this paper, we discuss the challenges of this project, and how the combined power of Biopython \cite{Biopython}, Matplotlib \cite{Matplotlib} and SWIG \cite{SWIG} were utilized for the required computational tasks. We finish by discussing how python goes beyond being a convenient programming language, and promotes good scientific practice by enabling clean code, integration with professional programming techniques such as unit testing, and strong data provenance.
\end{abstract}

\section{The Scientists Dilemma}

A typical scientific research project requires a variety of computational tasks to be performed.  At the very heart of every investigation is the generation of data to test hypotheses.  An experimental physicist builds instruments to collect light scattering data; a crystallographer collects X-ray diffraction data; a biologist collects fluorescence intensity data for reporter genes, or DNA sequence data for these genes; and a computational researcher writes programs to generate simulation data.  All of these scientists use computer programs to control instruments or perform simulations to collect and manage data in an electronic format.

Once data is collected, the next task is to analyze it in the context of hypothesis-driven models that help them understand the phenomenon they are studying.  In the case of light, or X-ray scattering data, there is a well-proven physical theory that is used to process the data and calculate the observed structure function of the material being studied \cite{Ashcroft-Mermin}.  This structure function is then compared to predictions made by the hypotheses begin tested.  In the case of biological reporter gene data, light intensity is matched up with phenotypic traits or DNA sequences, and statistically analyzed for trends that might explain the observed patterns.

As these examples illustrate, across science, the original raw data of each investigation is extensively processed by computational programs in an effort to understand the underlying phenomena.  Visualization tools to create a variety of scientific plots are often a preferred tool for both troubleshooting ongoing experiments, and creating publication-quality scientific plots and charts.  These plots and charts are often the final product of a scientific investigation in the form of data-rich graphics that demonstrate the truth of a hypothesis compared to its alternatives \cite{Tufte}.

Unfortunately, all too often scientists resort to a grab-bag of tools to perform these varied computational tasks.  For physicists and theoretical chemists, it is common to use C or FORTRAN to generate simulation data, and C code is used to control experimental apparatus; for biologists, perl is the language of choice to manipulate DNA sequence data \cite{Stein-bioperl}.  Data analysis is performed in separate, external software packages such as Matlab or Mathematica for equation solving \cite{Matlab,Mathematica}, or Stata, SPSS or R for statistical calculations \cite{Stata,SPSS,R}. Furthermore, separate data visualization packages can be used, making the scientific programming toolset extremely varied.

Such a mixed bag of tools is an inadequate solution for a variety of reasons.  From a computational perspective, most of these tools cannot be pipelined easily which necessitates many manual steps or excessive glue code that most scientists are not trained to write.  Far more important than just an inconvenience associated with gluing these tools together is the extreme burden placed on the scientist in terms of data management.  In complicated systems, there are often a plethora of different data files in several different formats residing at many different locations.  Most tools do not produce adequate metadata for these files, and scientists typically fall back on cryptic file naming schemes to indicate what type of data the files contain and how it was generated. Such complications can easily lead to mistakes.  This in turn provides poor at best data provenance when it is in fact of utmost importance in scientific studies where data integrity is the foundation of every conclusion reached and every fact established.  

Furthermore, when data files are manually moved around from tool to tool, it is not clear if an error is due to program error, or human error in using the wrong file.  Analyses can only be repeated by following work flows that have to be manually recorded in a paper or electronic lab notebook. This practice makes steps easily forgotten, and hard to pass on to future generations of scientists, or current peers trying to reproduce scientific results.

The Python programming language and associated community tools \cite{python} can help scientists overcome some of these problems by providing a general scientific programming platform that allows scientists to generate, analyze, visualize and manage their data within the same computational framework.  Python can be used to generate simulation data, or control instrumentation to capture data. Data analysis can be accomplished in the same way, and there are graphics libraries that can produce scientific charts and graphs.  Furthermore python code can be used to glue all of these python solutions together so that visualization code resides alongside the code that generates the data it is applied to.  This allows streamlined generation of data and its analysis, which makes data management feasible.  Most importantly, such a uniform tool set allows the scientist to record the steps used in data work flows to be written down in python code itself, allowing automatic provenance tracking.

In this paper, we outline a recent comparative genomics case study where python and associated community libraries were used as a complete scientific programming platform. We introduce several specific python libraries and tools, and how they were used to facilitate input of standardized biological data, create scientific plots, and provide solutions to speed bottle-necks in the code.  Throughout, we provide detailed tutorial-style examples of how these tools were used, and point to resources for further reading on these topics.  We conclude with ideas about how python promotes good scientific programing practices, and tips for scientists interested in learning more about python.

\section{A Comparative Genomics Case Study}

\begin{figure}
    \caption{Lambda phage GenBank file snippet.  The full file can be found online - see \cite{Lambda-GenBank}.}
    \label{lambda-genbank}
    {\tt \small
    \begin{verbatim}
LOCUS       NC_001416              48502 bp    DNA     linear   PHG 28-NOV-2007
DEFINITION  Enterobacteria phage lambda, complete genome.
ACCESSION   NC_001416
VERSION     NC_001416.1  GI:9626243
PROJECT     GenomeProject:14204
KEYWORDS    .
SOURCE      Enterobacteria phage lambda
  ORGANISM  Enterobacteria phage lambda
            Viruses; dsDNA viruses, no RNA stage; Caudovirales; Siphoviridae;
            Lambda-like viruses.
REFERENCE   1  (sites)
  AUTHORS   Chen,C.Y. and Richardson,J.P.
  TITLE     Sequence elements essential for rho-dependent transcription
            termination at lambda tR1
  JOURNAL   J. Biol. Chem. 262 (23), 11292-11299 (1987)
   PUBMED   3038914
...
FEATURES             Location/Qualifiers
     source          1..48502
                     /organism="Enterobacteria phage lambda"
                     /mol_type="genomic DNA"
                     /specific_host="Escherichia coli"
                     /db_xref="taxon:10710"
     gene            191..736
                     /gene="nu1"
                     /locus_tag="lambdap01"
                     /db_xref="GeneID:2703523"
     CDS             191..736
                     /gene="nu1"
                     /locus_tag="lambdap01"
                     /codon_start=1
                     /transl_table=11
                     /product="DNA packaging protein"
                     /protein_id="NP_040580.1"
                     /db_xref="GI:9626244"
                     /db_xref="GeneID:2703523"
                     /translation="MEVNKKQLADIFGASIRTIQNWQEQGMPVLRGGGKGNEVLYDSA
                     AVIKWYAERDAEIENEKLRREVEELRQASEADLQPGTIEYERHRLTRAQADAQELKNA
                     RDSAEVVETAFCTFVLSRIAGEIASILDGLPLSVQRRFPELENRHVDFLKRDIIKAMN
                     KAAALDELIPGLLSEYIEQSG"
...
ORIGIN      
        1 gggcggcgac ctcgcgggtt ttcgctattt atgaaaattt tccggtttaa ggcgtttccg
       61 ttcttcttcg tcataactta atgtttttat ttaaaatacc ctctgaaaag aaaggaaacg
      121 acaggtgctg aaagcgaggc tttttggcct ctgtcgtttc ctttctctgt ttttgtccgt
      181 ggaatgaaca atggaagtca acaaaaagca gctggctgac attttcggtg cgagtatccg
      241 taccattcag aactggcagg aacagggaat gcccgttctg cgaggcggtg gcaagggtaa
      301 tgaggtgctt tatgactctg ccgccgtcat aaaatggtat gccgaaaggg atgctgaaat
      361 tgagaacgaa aagctgcgcc gggaggttga agaactgcgg caggccagcg aggcagatct
      421 ccagccagga actattgagt acgaacgcca tcgacttacg cgtgcgcagg ccgacgcaca
...
\end{verbatim}}
\end{figure}

Recently we performed a comparative genomics study of the genomic DNA sequences of the 74 sequenced bacteriophages that infect \emph{E. coli}, \emph{P. aeruginosa}, or \emph{L. lactis} \cite{LucksJB-PlOSCompBio-2008}.  Bacteriophages are viruses that infect bacteria.  The DNA sequences of these bacteriophages contain important clues as to how the relationship with their host has shaped their evolution.  

Each virus that we examined has a DNA genome that is a long strand of four nucleotides called Adenine (A), Threonine (T), Cytosine (C), and Guanine (G).  The specific sequences of A's, T's, C's and G's encode for proteins that the virus uses to take over the host bacteria and create more copies of itself.  Each protein is encoded in a specific region of the genomic DNA called a gene.

Proteins are made up of linear strings of 20 amino acids.  There are 4 bases encoding for 20 amino acids, and the translation table that governs the encoding, called the genetic code, is comprised of 3 base triplets called codons.  Each codon encodes a specific amino acid.  Since there are 64 possible codons, and only 20 amino acids, there is a large degeneracy in the genetic code.  For more information on the genetic code, and the biological process of converting DNA sequences into proteins, see \cite{MolBiolCell}.

Because of this degeneracy, each protein can be `spelled' as a sequence of codons in many possible ways.  The particular sequence of codons used to spell a given protein in a gene is called the gene's `codon usage'.  As we found in \cite{LucksJB-PlOSCompBio-2008}, bacteriophages genomes favor certain codon spellings of genes over the other possibilites.  The primary question of our investigation was - does the observed spellings of the bacteriophage genome shed light onto the relationship between the bacteriophage and its host \cite{LucksJB-PlOSCompBio-2008}?

To address this question, we examined the codon usage of the protein coding genes in these bacteriophages for any non-random patterns compared to all the possible spellings, and performed statistical tests to associate these patterns with certain aspects about the proteins.

The computational requirements of this study included:

\begin{itemize}
    \item Downloading and parsing the genome files for viruses from GenBank in order to get the genomic DNA sequence, the gene regions and annotations: GenBank \cite{GenBank} is maintained by the National Center of Biotechnology Information (NCBI), and is a data wharehouse of freely available DNA sequences. For each virus, we needed to obtain the genomic DNA sequence, the parts of the genome that code for genes, and the annotated function of these genes.  Figure 1 displays this information for lambda phage, a well-studied bacteropphage that infects \emph{E. coli} \cite{MolBiolCell}, in GenBank format, obtained from NCBI.  Once these files were downloaded and stored, they were parsed for the required information.  
    \item Storing the genomic information: The parsed information was stored in a custom genome python class which also included methods for retrieving the DNA sequences of specific genes.
    \item Drawing random genomes to compare to the sequenced genome: For each genome, we drew random genomes according to the degeneracy rules of the genetic code so that each random genome would theoretically encode the same proteins as the sequenced genome.  These genomes were then visually compared to the sequenced genome through zero-mean cumulative sum plots discussed below.
    \item Visualize the comparisons through `genome landscape' plots: Genome landscapes are zero-mean cumulative sums, and are useful visual aids when comparing nucleotide frequency properties of the genomes they are constructed from (see \cite{LucksJB-PlOSCompBio-2008} for more information). Genome landscapes were computed for both the sequenced genome, and each drawn genome.  The genome landscape of the sequenced genome was compared to the distribution of genome landscapes generated from the random genomes to detect regions of the genomes that have extremely non-random patterns in codon usage.
    \item Statistically analyzing the non-random regions with annotation and host information: To understand the observed trends, we performed analysis of variance (ANOVA) \cite{ANOVA} analysis to detect correlations between protein function annotation or host lifestyle information with these regions.
    
\end{itemize}

Python was used in every aspect of this computational work flow.  Below we discuss in more detail how python was used in several of these areas specifically, and provide illustrative tutorial-style examples.  For more information on the details of the computational work flow, and the biological hypotheses we tested, see \cite{LucksJB-PlOSCompBio-2008}.  For specific details on the versions of software used in this paper, and links to free downloads, see Materials and Methods.

\section{Biopython}

Biopython is an open-source suite of bioinfomatics tools for the python language \cite{Biopython}.  The suite is comprehensive in scope, and offers python modules and routines to parse bio-database files, facilitate the computation of alignments between biological sequences (DNA and protein), interact with biological web-services such as those provided by NCBI, and examine protein crystallographic data to name a few.

In this project, Biopython was used both to download and parse genomic viral DNA sequence files from the NCBI Genbank database \cite{GenBank} as outlined in Listing \ref{genbank_code}.

\begin{lstlisting}[caption=Downloading and parsing the GenBank genome file for lambda phage (refseq number NC\_001416).,label=genbank_code]
# genbank.py - utilities for downloading and parsing GenBank files

from Bio import GenBank # (1)
from Bio import SeqIO

def download(accession_list):
    '''Download and save all GenBank records in accession_list.'''
    
    try:
        handle = GenBank.download_many(accession_list) # (2)
    except:
        print "Are you connected to the internet?"
        raise
    
    genbank_strings = handle.read().split('//\n') # (3)
    for i in range(len(accession_list)):  
        #Save raw file as .gb
        gb_file_name = accession_list[i]+'.gb'       
        f = open(gb_file_name,'w')
        f.write(genbank_strings[i]) # (4)
        f.write('//\n')
        f.close()


def parse(accession_list):
    '''Parse all records in accession_list.'''
    
    parsed = []
    for accession_number in accession_list:
        gb_file_name = accession_number+'.gb'
        print 'Parsing ... ',accession_number
        try:
            gb_file = file(gb_file_name,'r')
        except IOError:
            print 'Is the file %s downloaded?' % gb_file_name
            raise
        
        gb_parsed_record = SeqIO.parse(gb_file,"genbank").next() # (5)
        gb_file.close()
        
        print gb_parsed_record.id  # (6)
        print gb_parsed_record.seq
        
        parsed.append(gb_parsed_record) # (7)
    
    return parsed


import genbank # (8)
genbank.download(['NC_001416'])
genbank.parse(['NC_001416'])


# (1) The biopython module is called Bio.  The Bio.Genbank module is used to download records from GenBank, and the Bio.SeqIO module provides a general interface for parsing a variety of biological formats, including GenBank.

# (2) The Bio.GenBank.download\_many method is used in the genbank.download method to download Genbank records over the internet.  It takes a list of GenBank accession numbers identifying the records to be downloaded.

# (3) GenBank records are separated by the character string \verb=//\n=.  Here we manually separate GenBank files that are part of the same character string. 

# (4) When we save the GenBank records as individual files to disk, we include the \verb=//\n= separator again.

# (5) The Bio.SeqIO.parse method can parse a variety of formats.  Here we use it to parse the GenBank files on our local disk using the "genbank" format parameter.  The method returns a generator, who's next() method is used to retrieve an object representing the parsed file.

# (6) The object representing the parsed GenBank file has a variety of methods to extract the record id and sequence.  See Listing \ref{genome_code} for more details.

# (7) The genbank.parse method returns a listed of parsed objects, one for each input sequence file.

# (8) To run the code in genbank.py, Biopython 1.44 must first be installed (see Materials and Methods).  Executing the following code should create a file called `NC\_001416.gb' on the local disk (see Figure \ref{lambda-genbank}), as well as produce the following output:

'''
Parsing ...  NC_001416
NC_001416.1
Seq('GGGCGGCGACCTCGCGGGTTTTCGCTATTTATGAAAATTTTCCGGTTTAAGGCGTTTCCG ...', IUPACAmbiguousDNA())
'''

\end{lstlisting}

The benefits of using Biopython in this project were several including:
\begin{enumerate}
    \item Not having to write or maintain this code ourselves.  This is an important point as the number of web-available databases and services grows.  These often change rapidly, and require rigorous maintenance to keep up with tweaks to API's and formats - a monumental task that is completed by an international group of volunteers for the Biopython project.
    \item The Biopython parsing code can be wrapped in custom classes that make sense for a particular project. Listing \ref{genome_code} illustrates the latter by outlining a custom genome class used in this project to store the location of coding sequences for genes (CDS\_seq).
    
\end{enumerate}

\begin{lstlisting}[caption=A custom Genome class which wraps the biopython parsing code outlined in Listing \ref{genbank_code}.,label=genome_code]
# genome.py - a custom genome class which wraps biopython parsing code

import genbank # (1)
from Bio import Seq
from Bio.Alphabet import IUPAC

class Genome(object):
    """Genome - representing a genomic DNA sequence with genes
    
    Genome.genes[i] returns the CDS sequences for each gene i."""
    
    def __init__(self, accession_number):
        
        genbank.download([accession_number]) # (2)
        self.parsed_genbank = genbank.parse([accession_number])[0]
        
        self.genes = []
        
        self._parse_genes()
        
    
    def _parse_genes(self):
        """Parse out the CDS sequence for each gene."""
        
        for feature in self.parsed\_genbank.features: # (3)
            if feature.type == 'CDS':
                
                #Build up a list of (start,end) tuples that will
                #be used to slice the sequence in self.parsed\_genbank.seq
                #
                #Biopython locations are zero-based so can be directly
                #used in sequence splicing

                locations = []
                if len(feature.sub_features): # (4)
                    # If there are sub\_features, then this gene is made up
                    # of multiple parts.  Store the start and end positins
                    # for each part.
                    for sf in feature.sub_features:
                        locations.append((sf.location.start.position,
                                          sf.location.end.position))
                else:
                    # This gene is made up of one part.  Store its start and 
                    # end position.
                    locations.append((feature.location.start.position,
                                      feature.location.end.position))


                # Store the joined sequence and nucleotide indices forming
                # the CDS.
                seq = '' # (5)
                for begin,end in locations:
                    seq += self.parsed_genbank.seq[begin:end].tostring()

                # Reverse complement the sequence if the CDS is on
                # the minus strand  
                if feature.strand == -1:  # (6)
                  seq_obj = Seq.Seq(seq,IUPAC.ambiguous_dna)
                  seq = seq_obj.reverse_complement().tostring()

                # append the gene sequence
                self.genes.append(seq) # (7)

# (1) Here we import the genbank module outlined in Listing \ref{genbank_code}, along with two more biopython modules.  The Bio.Seq module has methods for creating DNA sequence objects used later in the code, and the Bio.Alphabet module contains definitions for the types of sequences to be used.  In particular we use the Bio.Alphabet.IUPAC definitions.

# (2) We use the genbank methods to download and parse the GenBank record for the input accession number.

# (3) The parsed object stores the different parts of the GenBank file as a list of features.  Each feature has a type, and in this case, we are looking for features with type 'CDS', which stores the coding sequence of a gene.

# (4) For many organisms, genes are not contiguous stretches of DNA, but rather are composed of several parts.  For GenBank files, this is indicated by a feature having sub\_features.  Here we gather the start and end positions of all sub features, and store them in a list of 2-tuples.  In the case that the gene is a contiguous piece of DNA, there is only one element in this list.

# (5) Once the start and end positions of each piece of the gene are obtained, we use them to slice the seq of the parsed\_genbank object, and collect the concatenated sequence into a string.

# (6) Since DNA has polarity, there is a difference between a gene that is encoded on the top, plus strand, and the bottom, minus strand.  The strand that the gene is encoded in is stored in feature.strand.  If the strand is the minus strand, we need to reverse compliment the sequence to get the actual coding sequence of the gene.  To do this we use the Bio.Seq module to first build a sequence, then use the reverse\_complement() method to return the reverse compliment.

# (7) We store each gene as an element of the Genome.genes list.  The CDS of the ith gene is then retrievable through Genome.genes[i].

\end{lstlisting}

For a more detailed introduction to the plethora of biopython features, as well as introductory information into python see \cite{Bassi-PLoSCompBio-2007} .

\section{Matplotlib}

Matplotlib \cite{Matplotlib} is a suite of open-source python modules that provide a framework for creating scientific plots similar to the Matlab \cite{Matlab} graphical tools. In this project, matplotlib was used to create genome landscape plots both to have a quick look at data as it was generated, and to produce publication quality figures.  Genome landscapes are cumulative sums of a zero-mean sequence of numbers, and are useful visualization tools for understanding the distribution of nucleotides across a genome (see \cite{LucksJB-PlOSCompBio-2008} for more information).

Listing \ref{landscape_code} outlines how matplotlib was used to quickly generate graphics to test raw simulation data as it was being generated.

\begin{lstlisting}[caption=Sample matplotlib script that calculates and plots the zero-mean cumulative sum of the numbers listed in a single column of an input file.,label=landscape_code]
# landscape.py - plotting a zero-mean cumulative sum of numbers

import fileinput # (1)
import numpy
from matplotlib import pylab

def plot(filename):
    """Read single-column numbers in filename and plot zero-mean cumulative sum"""
    
    numbers = []
    for line in fileinput.input(filename): # (2)
        numbers.append(float(line.split('\n')[0]))
    
    mean = numpy.mean(numbers) # (3)
    cumulative_sum = numpy.cumsum([number - mean for number in numbers])
    
    pylab.plot(cumulative_sum[0::10],'k-') # (4)
    pylab.xlabel('i')
    pylab.title('Zero Mean Cumulative Sum')
    
    pylab.savefig(filename+'.png') # (5)
    pylab.show()

# (1) We use several python community modules to plot the zero-mean cumulative sum.  As part of the python standard library, fileinput can be used as a quick an easy solution to reading in a file containing a column of entries.  numpy is a comprehensive python project aimed at providing numerical routines for scientific applications \cite{numpy}.  Finally we import the matplotlib.pylab module which provides a Matlab-like plotting environment.

# (2) Here we use fileinput to read successive lines of the input file, which takes care of opening and closing the input file automatically.  Notice that we split each line by the newline character \verb=\n=, and take everything to the left of it, assuming that each line contains a single number.

# (3) The numpy module provides many convenient methods such as mean to compute the \verb=mean= of a list of numbers, and \verb=cumsum= which computes the cumulative sum.  To shift the input numbers by the mean, we use a python list comprehension to subtract the mean from each number, and then input the shifted list to numpy.cumsum.

# (4) The pylab module presents a Matlab-like plotting environment.  Here we use several methods to create a basic line plot with an xlabel and title.

# (5) To view the plot, we use pylab.show(), after we have saved the figure as a PNG file using pylab.savefig. The following script uses the genome class outlined in Listing \ref{genome_code}, along with the landscape class to plot the GC-landscape for the lambda phage genome.  The genome class is used to download and parse the GenBank file for lambda phage.  Each gene sequence is then scanned for 'G' or 'C' nucleotides.  For every 'G' or 'C' nucleotide encountered, a 1 is appended to the list GC; for every 'A' or 'T' encountered, a 0 is appended.  This sequence of 1's and 0's representing the GC-content of the lambda phage genome is saved in a file, and input into the landscape.plot method.  A plot corresponding to executing this script is shown in Figure \ref{lambda-landscape}.

import genome,landscape
lambda_phage = genome.Genome('NC_001416')
GC = []
for gene_sequence in lambda_phage.genes:
    for nucleotide in gene_sequence:
        if nucleotide == 'G' or nucleotide == 'C':
            GC.append(1)
        else:
            GC.append(0)

f = file('NC_001416.GC','w')
for num in GC:
    f.write('%i\n' % num)
f.close()

landscape.plot('NC_001416.GC')

\end{lstlisting}

\begin{figure}
    \caption{The lambda phage GC-landscape generated by the sample code in Listing \ref{landscape_code}.\label{lambda-landscape}}
    \begin{center}
        \includegraphics[scale=0.6]{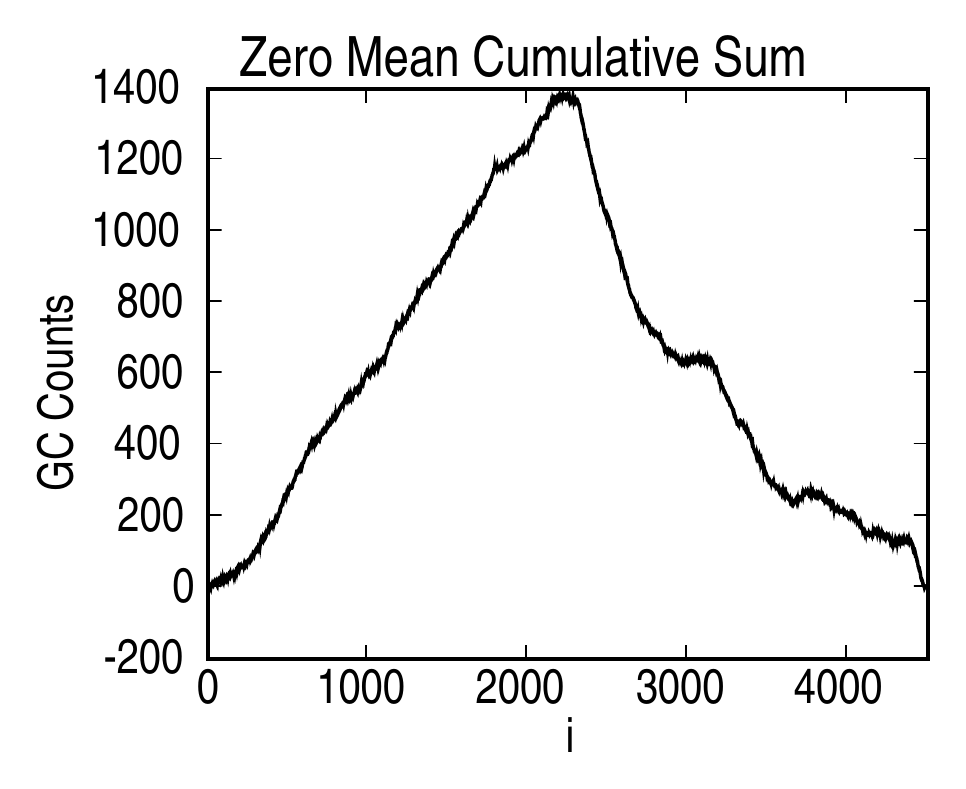}
    \end{center}
\end{figure}

Matplotlib was also used to make custom graphics classes for creating publication-quality plots.  To do this, we used the object oriented interface to matplotlib plotting routines to inherit funcionality in our classes.

The benefits of using matplotlib in this project were several:
\begin{enumerate}
    \item The code that produced the scientific plots resided alongside the code that produced the underlying data for the plots.  The importance of this cannot be stressed enough as having the code structured in this way removed many opportunities for human error involved in manually shuffling raw data files into separate graphical programs.  Moreover, the instructions for producing the plots from the underlying raw data was \emph{python code}, which not only described these instructions, but could be executed to produce the plots.  Imagine instead the often practiced use of spreadsheets to create plots from raw data - in these spreadsheets, formulas are hidden by the results of the calculations, and it is often very confusing to construct a picture of the computational flow used to produce a specific plot.
    \item Having the graphics instructions in code allowed for quick trouble shooting when creating the plots, or evaluating raw data as it was generated.
    \item Complicated plots were easily regenerated by tweaking the code for particular graphical plots.
    
\end{enumerate}

\section{SWIG}

The Simple Wrapper and Interface Generator (SWIG) \cite{SWIG}, is an easy-to-use system for extending python.  In particular, it allows the speed up of selected parts of an application by writing these routines in another more low-level language such as C or C++.  Furthermore, SWIG implements the use of this low-level code using the standard python module importing structure.  This allows developers to first prototype code in python, then re-implement the code in C and SWIG causing \emph{no change} in the python code that uses the re-implemented module.

This project relied heavily on drawing random numbers from an input discrete distribution.  For example, we often needed to draw a sequence of A's, T's, C's or G's corresponding to the nucleotide sequence of the genome, but preserving the genomic distribution of these four nucleotide bases.  For some viruses, the distribution might look like: $P_A = 0.2$, $P_T = 0.2$, $ P_C = 0.3 $, $ P_G = 0.3 $, with $P_A + P_T + P_C + P_G = 1.0 $.  Listing \ref{random_python_code} illustrates the outline of a python module that has methods to draw numbers according to a discrete distribution with 4 possible outcomes.  It also illustrates how this module could be implemented in C, and included in a python module with SWIG.

\begin{lstlisting}[caption=Drawing random numbers from a specified discrete distribution with four possibilities implemented in python. ,label=random_python_code]
# module discrete\_distribution.py - drawing numbers from a discrete probability distribution

import random # (1)

def seed(): # (2)
    random.seed()
    
def draw(distribution): # (3)
  '''Drawing an index according to distribution.
  
  distribution is a list of floating point numbers,
  one for each index number, representing the probability 
  of drawing that index number.
  
  Example: [0.5, 0.5] would represent equal probabilities
  of returning a 0 or 1.
  '''
  sum = 0 # (4)
  r = random.random()
  for i in range(0,len(distribution)):
      sum += distribution[i]
      if r < sum:
          return i


import discrete_distribution # (5)
discrete_distribution.seed()
print sum([discrete_distribution.draw([0.2,0.2,0.3,0.3]) for x in range(10000)])/10000. 

# (1) Import the random number generator.

# (2) We use the discrete\_distribution.seed() method to seed the random number generator.  If no arguments are supplied to random.seed(), the system time is used to seed the number generator \cite{python-random}.

# (3) The draw function takes an argument distribution, which is a list of floating point numbers.

# (4) The algorithm for drawing a number according to a discrete distribution is to draw a number, r, from a uniform distribution on [0,1]; compute a cumulative sum of the probabilities in the discrete distribution for successive indices of the distribution; when r is less than this cumulative sum, return the index that the cumulative sum is at.

# (5) To test this code, plug in a distribution [0.2,0.2,0.3,0.3], draw 10000 numbers from this distribution, and compute the mean, which theoretically should be $0*0.2 + 1*0.2 + 2*0.3 + 3*0.3 = 1.7$.  In this case, when this code was executed, the result $1.7013$ was returned.

\end{lstlisting}

In Listing \ref{random_C_code}, we implement this routine using C, and use SWIG to create a python module of the C implementation.

\lstset{language=C}
\begin{lstlisting}[caption=Drawing random numbers from a specified discrete distribution with four possibilities implemented in C with SWIG. ,label=random_C_code]

//c\_discrete\_distribution.c - A C implementation of the discrete\_distribution.py module

#include "stdlib.h" // (1)
#include "stdio.h"
#include "time.h"

void seed() {
    srand((unsigned) time(NULL) * getpid()); 
}

int draw(float distribution[4]) { // (2)
    float r= ((float) rand() / (float) RAND_MAX);
    float sum = 0.;
    int i = 0;
    for(i = 0; i < 4; i++) {
        sum += distribution[i];
  if (r lt sum) {
            return i;
  }
    }
}

// (1) Here we define two functions, seed and draw, which correspond to the python methods in discrete\_distribution.py.  Note that the python implementation of discrete\_distribution.draw() worked with distributions of arbitrary numbers of elements.  For simplicity, we are restricting the C implementation to work with distributions of length 4.

// (2) The draw routine is implemented using the same algorithm as in the python implementation.  For simplicity, we use the C standard library rand() routine, although there are more advanced random number generators that would be more appropriate for scientific applications \cite{random}. (Note that the `lt' symbol should be replaced by `<' when executing the code.)

// c\_discrete\_distribution.i - A Swig interface file for the c\_discrete\_distribution module // (3)

%module c_discrete_distribution // (4)

// Grab a 4 element array as a Python 4-list // (5)
%typemap(in) float[4](float temp[4]) {   // temp[4] becomes a local variable
  int i;
  if (PyList_Check($input)) {
    PyObject* input_to_tuple = PyList_AsTuple($input);
    if (!PyArg_ParseTuple(input_to_tuple,"ffff",temp,temp+1,temp+2,temp+3)) {
      PyErr_SetString(PyExc_TypeError,"tuple must have 4 elements");
      return NULL;
    }
    $1 = &temp[0];
  } else {
    PyErr_SetString(PyExc_TypeError,"expected a tuple.");
    return NULL;
  }
}

void seed(); // (6)
int draw(float distribution[4]); 


// (3) To use SWIG, we create a swig interface file that describes how to translate python inputs to the C code, and C outputs to the python code.

// (4) SWIG directives are preceded by the \% sign.  Here we declare that the module we are going to make is called c\_discrete\_distribution.  In general, the module name, the C source name, and the interface file name should all be the same outside of the file extension.

// (5) SWIG will automatically handle the conversion of many data-types from python to C and C to python.  For illustration purposes, we create an explicit typemap which converts a 4-element python list into a 4 element C list of floats.  Since we are using the typemap(in) directive, SWIG knows that we are converting python to C.  The rest of the code checks that a list was passed from python to C, and the list has 4 elements.  If these conditions are not met, python errors are thrown.  If they are met, an array of floats called temp is called, and passed to C.  This conversion is adapted from the SWIG reference manual \cite{SWIG}.

// (6) The last thing to do in the SWIG interface file is to declare the function signatures of the C implementation.

\end{lstlisting}

To use this module outlined in Listing \ref{random_C_code}, we have to call swig to generate wrapper code, then compile and link our code with the wrapper code.  With SWIG installed, the procedure would look something like

\begin{verbatim}
swig -python -o c_discrete_distribution_wrap.c c_discrete_distribution.i
\end{verbatim}

We first use SWIG to generate the wrapper code.  Using the c\_discrete\_distribution.i interface file, SWIG will generate c\_discrete\_distribution\_wrap.c using the Python C API, since we specified the -python flag.  In addition, SWIG will also generate c\_discrete\_distribution.py, which we will use to import the module into our code.

\begin{verbatim}
gcc -c c_discrete_distribution.c c_discrete_distribution_wrap.c
    -I/usr/include/python2.5 -I/usr/lib/python2.5
\end{verbatim}

Next we use a C compiler to compile each of the C files (our C source, and the SWIG generated wrapper).  We have to include the python header files and libraries for the python version we are using.  In our case, we used python 2.5.  After this procedure completes, we should have two additional files: c\_discrete\_distribution.o and c\_discrete\_distribution\_wrap.o .

\begin{verbatim}
gcc -bundle -flat_namespace -undefined suppress 
    -o _c_discrete_distribution.so 
       c_discrete_distribution.o c_discrete_distribution_wrap.o
\end{verbatim}

The final step is to link them all together.  The linking options are platform dependent, and the official SWIG documentation should be consulted \cite{SWIG}. For Mac OS X, we use the ``-bundle -flat\_namespace -undefined suppress'' options for gcc.  When this step is done, the file \_c\_discrete\_distribution.so is created.

The python module file c\_discrete\_distribution.py can be used in the same way as in Listing \ref{random_python_code} above, 

\begin{verbatim}
import c_discrete_distribution as discrete_distribution
discrete_distribution.seed()
print sum([discrete_distribution.draw([0.2,0.2,0.3,0.3]) for x in range(10000)])/10000. 
\end{verbatim}

\noindent which produces the number 1.6942.

The benefits of using SWIG in this project were several:
\begin{enumerate}
    \item We used all the benefits of python with the increased speed for critical bottlenecks of our simulation code.
    \item The parts that were sped up were used in the exact same context through the python module import structure, removing the need for glue code to tie in external C-programs.
    
\end{enumerate}

More generally, SWIG allows scientists using python to leverage experience in other languages that they typically have, while staying within the python framework with all its benefits outlined above.  This promotes a scientific work flow which consists of prototyping simulation code using the more simple python, then profiling the python code to identify the speed bottlenecks.  These can then be re-implemented in C or C++ and wrapped into the existing python code using SWIG.  This is a much preferred methodology than writing unnecessarily complicated and error-prone C programs, and using glue code to integrate them within the larger simulation methodology.

\section{Conclusions}

There are several practical conclusions to draw for scientists.  The first is that python, and its associated modules supported by the python community, offer a general platform for computing that is useful accross a broad range of scientific disciplines.  We have only outlined several such tools in this article, but there exist many more relevant to scientists \cite{scipy}.  The second is that python and its community modules can \emph{easily} be used by scientists.  The clean nature of the code is quick to learn, and its high-level features make complicated tasks quick to accomplish.  We have not discussed the interactive programming environments offered by python\cite{ipython,perez-ipython}, which when combined with the power of the language makes prototyping ideas and algorithms extremely easy.

The bigger picture conclusion is that python promotes good scientific practice.  The code readability and package structure enables code to be easily understood by different researchers working on the same project.  In fact, python code is often self-documenting which allows researchers to go back to code they wrote in the past and easily understand it.  Python and its community modules provide a consistent framework to generate data, and shuttle it to the various analysis tasks.  This in turn promotes data provenance through a written record \emph{in code} of every step used to analyze specific data, which removes many manual steps, and thus many errors.  

Finally, by using python, scientists can start to use other community tools and practices originally designed for professional programmers, but also useful to scientists.  The most important of these, but not discussed in this article, is unit testing, whereby test code is written alongside scientific code that tests to see if that code is working properly.  This allows scientists to re-write aspects of the code, perhaps using a different algorithm, and to re-run the tests to see if it still works as they think it should.  For large projects this is critical, and removes the need for often-used ad-hoc practices of looking at some sample data by eye, which is not only tedious, but not guaranteed to uncover subtle numerical bugs that could cause crucial mis-interpretation of scientific data.

Since python is a well-established language and has a large and active community, the resources available for beginners can be overwhelming.  For the scientist interested in learning more about scientific programming in python, we recommend visiting the web page and mailing lists of the SciPy project for an introduction to scientific modules \cite{scipy}, and \cite{byte-of-python,dive-into-python} for excellent introductory python tutorials.

\section{Materials and Methods}

All code examples in this paper were written by the author.  The particular versions of the relevant software used were: Python 2.5, Biopython 1.44, MatPlotLib 0.91.2, and SWIG 1.3.33.  Documentation and free downloads of this software are available at the following URLs:
\begin{itemize}
    \item Python - \href{http://python.org}{http://python.org}
    \item Biopython - \href{http://biopython.org}{http://biopython.org}
    \item MatPlotLib - \href{http://matplotlib.sourceforge.net}{http://matplotlib.sourceforge.net}
    \item SWIG - \href{http://www.swig.org/}{http://www.swig.org/}
\end{itemize}

The source code for all the listings above, as well as the original and maintained version of this article can be found at \verb=http://openwetware.org/wiki/Julius_B._Lucks/Projects/Python_All_A_Scientist_Needs=.
    
\section{Acknowledgements}

The author would like to thank Adrian Del Maestro, Joao Xavier, David Thompson and Stanley Qi for helpful comments during the preparation of this manuscript.  The author also thanks the Miller Institute for Basic Research in Science at the University of California, Berkeley for support.

\section{References and Resources}

\end{document}